\documentclass[aps,prl,twocolumn,floats,letterpaper,superscriptaddress,tighten]{revtex4}
\UseRawInputEncoding

\usepackage{amssymb}
\usepackage{amsbsy}
\usepackage{amsmath}
\usepackage{graphicx}
\usepackage{graphics}
\usepackage{setspace}
\usepackage{array}
\usepackage{color}
\usepackage{xcolor}
\usepackage{fontenc}
\usepackage{textcomp}
\usepackage{rotating}
\usepackage{bm}
\usepackage{hyperref}
\hypersetup{pdfpagemode=UseNone}

\newcommand{\e}{\varepsilon}

\newcommand{\vex}[1]{\bm{\mathrm{#1}}}

\newcommand{\bsub}{\begin{subequations}}
\newcommand{\esub}{\end{subequations}}

\newcommand{\be}{\begin{equation}}
\newcommand{\ee}{\end{equation}}
\newcommand{\bea}{\begin{eqnarray}}
\newcommand{\eea}{\end{eqnarray}}

\begin{document}
\title{The kinetic theory of ultra-subsonic fermion systems and applications to flat band magic angle twisted bilayer graphene}
\author{Seth M.\ Davis}
\email{smdavis1@umd.edu}
\affiliation{Condensed Matter Theory Center and Joint Quantum Institute, Department of Physics, University of Maryland, College Park, MD 20742, USA}
\author{Sankar\ Das\ Sarma}
\affiliation{Condensed Matter Theory Center and Joint Quantum Institute, Department of Physics, University of Maryland, College Park, MD 20742, USA}
\date{\today}

%----------------------------------------------------------------------------------
%%%%%%%%%%%%%%%%%%%%%%%%%%%%%%%%%%%%%%%%%%%%%%%%%%%%%%%%%%%%%%%
\begin{abstract}
The only kinematically-allowed phonon-scattering events for bands of subsonic fermions ($v_F < v_p$) are interband transitions, leading to different low-$T$ transport physics than in the typical supersonic case. We apply a kinetic theory of phonon-limited transport to a generic two-band system of subsonic fermions, deriving formulae for relaxation times and resistivity that are accurate in the limit of close bands and small $v_F/v_p$. We predict regimes of $\rho \propto T$, $\rho \propto T^2$, and perfect conductivity. Our theory predicts linear-in-$T$ resistivity down to a crossover temperature that is suppressed from its supersonic analogue by a factor of $v_F/v_p$, offering a new explanation for low-$T$ ``strange metal" behavior observed in flat band systems.
\end{abstract}
\maketitle

%%%%%%%%%%%%%%%%%%%%%%%%%%%%%%%%%%%%%%%%%%%%%%%%%%%%%%%%%
%%%%%%%%%%%%%%%%%%%%%%%%%%%%%%%%%%%%%%%%%%%%%%%%%%%%%%%%%
%%%%%%%%%%%%%%%%%%%%%%%%%%%%%%%%%%%%%%%%%%%%%%%%%%%%%%%%%
%%%%%%%%%%%%%%%%%%%%%%%%%%%%%%%%%%%%%%%%%%%%%%%%%%%%%%%%%
%%%%%%%%%%%%%%%%%%%%%%%%%%%%%%%%%%%%%%%%%%%%%%%%%%%%%%%%%
%%%%%%%%%%%%%%%%%%%%%%%%%%%%%%%%%%%%%%%%%%%%%%%%%%%%%%%%%
%%%%%%%%%%%%%%%%%%%%%%%%%%%%%%%%%%%%%%%%%%%%%%%%%%%%%%%%%

% Introduction

%%%%%%%%%%%%%%%%%%%%%%%%%%%%%%%%%%%%%%%%%%%%%%%%%%%%%%%%%
%%%%%%%%%%%%%%%%%%%%%%%%%%%%%%%%%%%%%%%%%%%%%%%%%%%%%%%%%
%%%%%%%%%%%%%%%%%%%%%%%%%%%%%%%%%%%%%%%%%%%%%%%%%%%%%%%%%
%%%%%%%%%%%%%%%%%%%%%%%%%%%%%%%%%%%%%%%%%%%%%%%%%%%%%%%%%
%%%%%%%%%%%%%%%%%%%%%%%%%%%%%%%%%%%%%%%%%%%%%%%%%%%%%%%%%
%%%%%%%%%%%%%%%%%%%%%%%%%%%%%%%%%%%%%%%%%%%%%%%%%%%%%%%%%
%%%%%%%%%%%%%%%%%%%%%%%%%%%%%%%%%%%%%%%%%%%%%%%%%%%%%%%%%

Rapid progress in the fabrication and manipulation of layered two-dimensional van der Waals heterostructures has lead to an unprecedented ability to engineer nearly flat band (NFB) electronic systems, which have already displayed a wealth of exotic phenomena \cite{Geim_2013, Novoselov_2006, Bistritzer_2011, Morell_2010, Li_2019, Kim_2017, Cao_2018a, Cao_2018b, Cao2020PRL, Cao2021, Yankowitz_2019,Kerelsky_2019, Lu_2019, Stepanov2020untying, Sharpe_2019, Chen_2020, Rozen2021entropic, AndreaYoungBernal, AndreaYoungRhombo, AndreaYoungRhombo2, Serlin_2020, Wu_2018, Wu_2019_TIPRL, KinFaiMak_TopologyTMD, Polshyn_2020, TBLGStrangeMetalExperiment1, Polshyn2019, TBLGStrangeMetalExperiment3, SankarFengchengStrangeMetal, CalTechBernal, CalTechSymmetryBreakingTBLG, Xie_2020, MacDonaldTLBGReview, Li_2021, Ghiotto_2021, Pan_2020, Pan_2021, Morales_Dur_n_2021, Ahn_2022, Kerelsky2021moireless, Khalaf2019, CalTechTwisted}. However, all solid state systems contain phonons. To understand observations of novel physics in solid state NFB systems, it is important to understand how phonons interact with NFB fermions. This work focuses on one aspect: when the fermions in question are \textit{subsonic} ($v_F < v_p$), kinematics requires that all single-phonon scattering processes are interband transitions, with consequences on the low-$T$ transport physics.

The prime example of the NFB systems is magic angle twisted bilayer graphene (MATBLG). MATBLG has been found to exhibit SC proximate to strongly correlated insulating states \cite{Cao_2018a, Cao_2018b, Cao2020PRL, Cao2021, Yankowitz_2019, Kerelsky_2019, Lu_2019, Wu_2018} and has been reported to exhibit a linear-in-$T$ ``strange metal"-like resistivity over a large range of dopings and temperatures \cite{TBLGStrangeMetalExperiment1, Polshyn2019, TBLGStrangeMetalExperiment3, SankarFengchengStrangeMetal}, sometimes down to temperatures as low as $50mK$ \cite{Jaoui_2022}. These phenomena have inspired analogy between MATBLG and the cuprate high-$T_c$ SCs, as well as speculation that SC in MATBLG might be driven by strong correlation physics. However, phonon-based theories of SC \cite{Wu2019_phonon, Li2020} and high-$T$ transport \cite{Wu2019_phonon, Davis_2023, SankarFengchengStrangeMetal} in MATBLG have been put forth that give generally good quantitative agreement with experiment. These results - along with many others \cite{Stepanov2020untying, Saito2020independent} - suggest an alternative physical picture in which MATBLG hosts standard, phonon-driven BCS-style SC that competes with interaction-driven insulating orders at commensurate filling fractions of the NFBs \cite{Lu_2019}. It is thus imperative to understand whether low-$T$, linear-in-$T$ transport in MATBLG is indeed arising from a strange metal state.

The standard kinetic theory of acoustic phonon scattering is exceptionally accurate in describing transport in layered graphene systems (as well as in normal metals and semiconductors) at temperatures above a few Kelvin \cite{Hwang2008,Wu2019_phonon,Davis_2022, Davis_2023}, and accurately describes TBLG transport away from the magic angle \cite{Wu2019_phonon,Davis_2023}. In this work, we apply the same framework to NFB systems at asymptotically low $T$, extending the kinetic theory well beyond its regime of proven validity. Remarkably, the familiar theory predicts qualitatively different low-$T$ transport physics for subsonic fermions than it does for the supersonic alternative, due entirely to the kinematic differences between the two limits. 

We develop a transport theory for an ``ultrasubsonic" (USS) two-band fermion system, which is accurate in the asymptotic double-limit of small $v_F/v_p$ and small band separation. Our main results are analytical formulae for relaxation times and for the resistivity of a general USS system, with transparent dependencies on temperature and doping level. We show that the interband nature of subsonic fermion scattering manifests an exponentially divergent relaxation time at low temperatures, in contrast with the familiar $\tau \propto T^{-4}$ Bloch-Gr\"{u}neisen (BG) power law applicable to supersonic  fermions. The divergence in relaxation time is capable of perfectly balancing the thermodynamic suppression of states away from the Fermi level, leading to a non-intuitive physical picture in which states far from the Fermi level contribute meaningfully to transport. This scenario predicts a linear-in-$T$ resistivity over a wide range of $T$, and down to a temperature much lower than the BG paradigm of supersonic fermion bands would suggest possible. The divergence in relaxation time can also drive the system to perfect conductivity at asymptotically low $T$, mimicking a SC transition. We also note that the geometry of isolated NFBs can manifest a mid-$T$ $\rho \propto T^2$ power law. All these features are consistent with the hitherto unexplained phenomenology of MATBLG transport.

\begin{figure}[t!]
\includegraphics[angle=0,width=.47\textwidth]{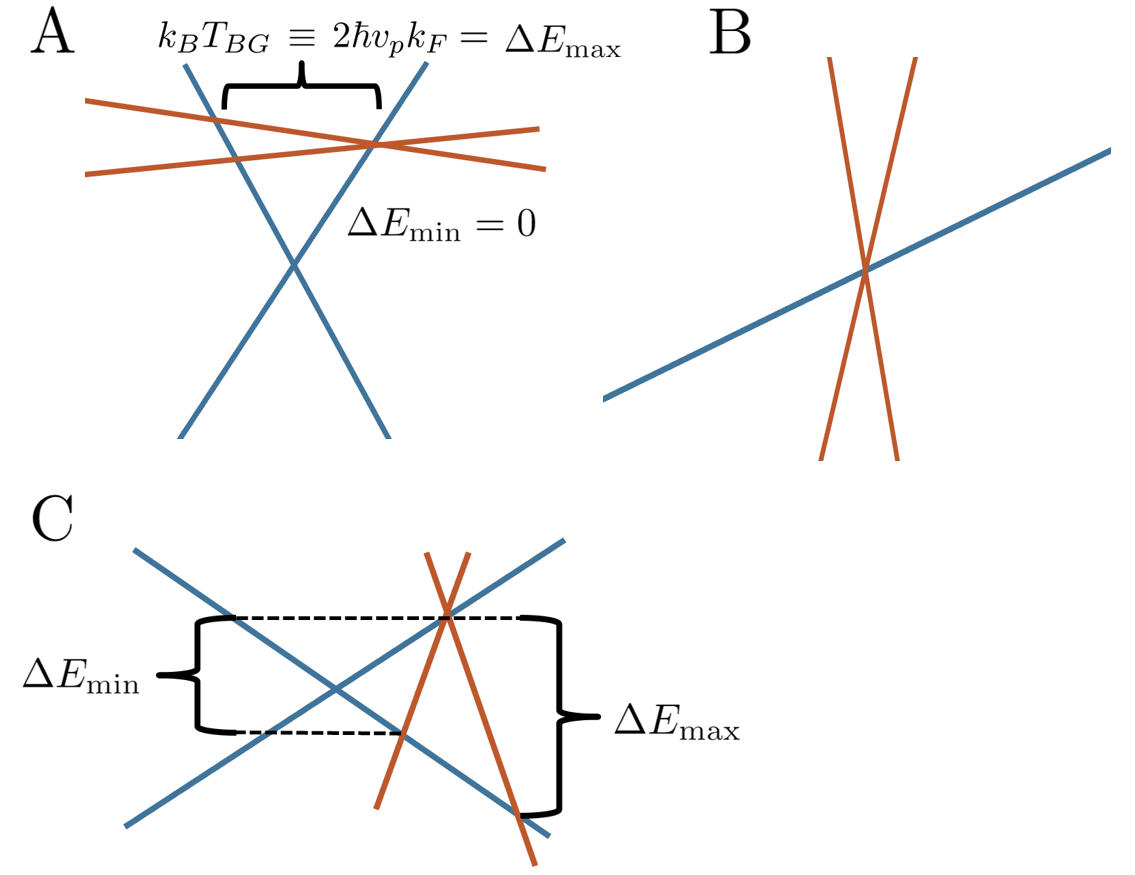}
\caption{Key aspects of phonon scattering for subsonic fermions can be understood from simple schematic drawings. We consider a cross section of a fermion band structure (blue), with momentum on the horizontal axis and energy on the vertical axis. We pick an ``initial" Bloch state and superimpose on it the Debye dispersion of the acoustic phonons (orange). Due to energy and momentum conservation, a fermion in the initial state can only scatter to a new state at an intersection point of the fermion and phonon bands. In (A), we depict the standard case of phonon scattering for supersonic (Dirac) fermions. The maximum allowed energy transfer defines $T_{BG}$, while arbitrarily small energy transfers are allowed. In (B), we show that a single-band system of subsonic fermions does not allow single-phonon scattering events. This follows from the simple fact that the fermion and phonon dispersions only intersect at the initial state. In (C) we consider a multiband subsonic fermion system, where phonon scattering is necessarily interband. In addition to a maximum allowed energy transfer defining a Bloch-Gr\"{u}neisen scale, there is also a minimum allowed energy transfer.}
\label{SubsonicSchematic1}
\end{figure}

There are reasons to suspect that the standard kinetic theory of acoustic-phonon-limited-transport may not apply to NFB systems. Kinetic theory replaces the interacting Hamiltonian with a Markovian collision integral, and is thus incapable of treating emergent physics due to strong correlations, which many expect to be important in NFB systems. Further, vertex corrections to the electron-phonon coupling could become important in the absence of Migdal's theorem. Nevertheless, it is important to understand the predictions of naive kinetic theory for NBF systems. The fact that our theory gives a simple and concrete mechanism for robust linear-in-$T$ resistivity is noteworthy, given that this is the primary signature of the ``strange metal" state. Further, it is interesting that phonon scattering in subsonic fermion bands can also generate regimes of $\rho \propto T^2$ resistivity - often assumed to arise from electron-electron scattering in Fermi liquids - and regimes of perfect conductivity that mimic the behavior of superconductivity. While a realistic treatment of NFB electron-phonon physics is beyond reach, it is possible that the atypical kinematics of subsonic fermions is a key ingredient to understanding transport in NFB systems, which our approach is able to fully capture. 

Although our theory has MATBLG in mind, we work with a general USS model. Our transport theory could apply to a system with renormalized fermion and phonon band structures and a renormalized deformation potential coupling in a theory that has already taken interaction-induced corrections into account. This, in fact, is the standard transport theory in normal metals where electron-electron interactions generally do not affect transport in the phonon-dominated regime. Further, in addition to providing an alternative explanation for experimental transport observations in MATBLG, our results could find application generally in all subsonic band systems, including other flat band systems generated by layered 2D heterostructures and heavy fermion systems. It is also noteworthy that strange metallicity is ubiquitous in heavy fermion compounds, which also manifest flat bands (i.e. heavy mass) and multiple bands near the Fermi surface, indicating a possible role for our proposed subsonic transport mechanism.

%%%%%%%%%%%%%%%%%%%%%%%%%%%%%%%%%%%%%%%%%%%%%%%%%%%%%%%%%
%%%%%%%%%%%%%%%%%%%%%%%%%%%%%%%%%%%%%%%%%%%%%%%%%%%%%%%%%
%%%%%%%%%%%%%%%%%%%%%%%%%%%%%%%%%%%%%%%%%%%%%%%%%%%%%%%%%
%%%%%%%%%%%%%%%%%%%%%%%%%%%%%%%%%%%%%%%%%%%%%%%%%%%%%%%%%
%%%%%%%%%%%%%%%%%%%%%%%%%%%%%%%%%%%%%%%%%%%%%%%%%%%%%%%%%
%%%%%%%%%%%%%%%%%%%%%%%%%%%%%%%%%%%%%%%%%%%%%%%%%%%%%%%%%
%%%%%%%%%%%%%%%%%%%%%%%%%%%%%%%%%%%%%%%%%%%%%%%%%%%%%%%%%

% Subsonic fermion scattering

%%%%%%%%%%%%%%%%%%%%%%%%%%%%%%%%%%%%%%%%%%%%%%%%%%%%%%%%%
%%%%%%%%%%%%%%%%%%%%%%%%%%%%%%%%%%%%%%%%%%%%%%%%%%%%%%%%%
%%%%%%%%%%%%%%%%%%%%%%%%%%%%%%%%%%%%%%%%%%%%%%%%%%%%%%%%%
%%%%%%%%%%%%%%%%%%%%%%%%%%%%%%%%%%%%%%%%%%%%%%%%%%%%%%%%%
%%%%%%%%%%%%%%%%%%%%%%%%%%%%%%%%%%%%%%%%%%%%%%%%%%%%%%%%%
%%%%%%%%%%%%%%%%%%%%%%%%%%%%%%%%%%%%%%%%%%%%%%%%%%%%%%%%%
%%%%%%%%%%%%%%%%%%%%%%%%%%%%%%%%%%%%%%%%%%%%%%%%%%%%%%%%%

\textit{Phonon scattering of supersonic and subsonic fermions}.---In an electron-phonon scattering process, conservation of energy and momentum define a ``scattering manifold" of kinematically allowed Bloch states that an electron can scatter to. The maximum energy difference between a Bloch state on the scattering manifold and the initial state defines the Bloch-Gr\"{u}neisen temperature, $T_{BG}$ \cite{Hwang_2019}. For example, in an isotropic system, we simply have $k_B T_{BG} \equiv 2 v_p k_F$ [Fig.\ref{SubsonicSchematic1} (A)]. When $T \gg T_{BG}$, enough phonon modes will be populated that all kinematically allowed scattering events are possible. This is the so-called ``equipartition regime", which is characterized by a linear-in-$T$ scattering rate for each Bloch state. This is universal in kinetic theory and usually (but not always \cite{Davis_2022, Davis_2023}) gives a linear-in-$T$ resistivity above $T_{BG}$. On the other hand, when $T \ll T_{BG}$, only low-energy phonon modes are available and scattering is restricted to a small neighborhood of the initial state. This is what gives rise to the famous low-$T$ $\tau \propto T^{-(d+2)}$ power law in the Bloch-Gr\"{u}neisen regime (where $d$ is the spatial dimension). Crucially, when the electrons are supersonic, the scattering manifold is smoothly connected to the original state, allowing arbitrarily small-momentum scattering events. As a result, the BG regime holds all the way to zero temperature. In single-layer graphene and in normal metals, $v_F/v_p \approx \mathcal{O}(10^2)$ or larger. (We also note that in many normal metals, the Debye temperature is lower than $T_{BG}$, in which case the BG crossover takes place at the Debye temperature \cite{Hwang_2019}.)

However, when the Bloch state in question is subsonic, the scattering manifold determined by energy-momentum conservation is necessarily disconnected from the original state. Similarly, in a band made entirely of subsonic Bloch states, all phonon scattering processes in subsonic fermion systems are interband processes. This implies that the scattering manifold also has a minimum allowed energy transfer [see Fig.~\ref{SubsonicSchematic1} (C)], defining another temperature scale applicable for subsonic scattering, $T_{SS}.$ Below $T_{SS}$ we expect an crossover to a regime in which the entire scattering manifold is inaccessible. That is, all thermally populated phonons are too low-energy to satisfy the conservation laws of an electron-phonon scattering event. Intuitively, we expect that below $T_{SS}$, the relaxation time of the Bloch state diverges exponentially as $T \rightarrow 0$ (instead of $\tau \propto T^{-4}$ as in the supersonic BG regime).

%%%%%%%%%%%%%%%%%%%%%%%%%%%%%%%%%%%%%%%%%%%%%%%%%%%%%%%%%
%%%%%%%%%%%%%%%%%%%%%%%%%%%%%%%%%%%%%%%%%%%%%%%%%%%%%%%%%
%%%%%%%%%%%%%%%%%%%%%%%%%%%%%%%%%%%%%%%%%%%%%%%%%%%%%%%%%
%%%%%%%%%%%%%%%%%%%%%%%%%%%%%%%%%%%%%%%%%%%%%%%%%%%%%%%%%
%%%%%%%%%%%%%%%%%%%%%%%%%%%%%%%%%%%%%%%%%%%%%%%%%%%%%%%%%
%%%%%%%%%%%%%%%%%%%%%%%%%%%%%%%%%%%%%%%%%%%%%%%%%%%%%%%%%
%%%%%%%%%%%%%%%%%%%%%%%%%%%%%%%%%%%%%%%%%%%%%%%%%%%%%%%%%

% Basic kinetic theory

%%%%%%%%%%%%%%%%%%%%%%%%%%%%%%%%%%%%%%%%%%%%%%%%%%%%%%%%%
%%%%%%%%%%%%%%%%%%%%%%%%%%%%%%%%%%%%%%%%%%%%%%%%%%%%%%%%%
%%%%%%%%%%%%%%%%%%%%%%%%%%%%%%%%%%%%%%%%%%%%%%%%%%%%%%%%%
%%%%%%%%%%%%%%%%%%%%%%%%%%%%%%%%%%%%%%%%%%%%%%%%%%%%%%%%%
%%%%%%%%%%%%%%%%%%%%%%%%%%%%%%%%%%%%%%%%%%%%%%%%%%%%%%%%%
%%%%%%%%%%%%%%%%%%%%%%%%%%%%%%%%%%%%%%%%%%%%%%%%%%%%%%%%%
%%%%%%%%%%%%%%%%%%%%%%%%%%%%%%%%%%%%%%%%%%%%%%%%%%%%%%%%%

\textit{Basic kinetic theory}.---We first review the fundamental equations for phonon-limited resistivity in the Boltzmann framework \cite{Davis_2022, Davis_2023}. The resistivity is given in terms of the \textit{relaxation times} $\tau_S$:
\begin{align}
\label{ResistivityDefinition}
    \frac{\delta^{ij}}{\rho}
    &=
    \frac{1}{4}
    \frac{e^2}{k_B T}
    \frac{1}{\mathcal{A}}
    \sum_{S}
    \frac{
    \tau_{S}
    v^i_{S}v^j_{S}
    }
    {
    \cosh\left(
    \frac{\e-\mu}{2 k_B T}
    \right)^2
    },
\end{align}
where $\mathcal{A}$ is the area of the system and $v_S$ is the Fermi velocity of the state $S$. The relaxation times are determined from a self-consistent integral equation, derived from the Boltzmann equation. In the case of longitudinal acoustic phonons, in the Debye approximation, which couple to the fermions via the deformation potential, this takes the form
\begin{align}
    \label{RelaxationLengthSelfConsistency}
    \frac{\pi D^2}{\hbar \rho_M v_p^2}
    \frac{1}{\mathcal{A}}
    \sum_{S'}
    \tilde{\Delta}_{S,S'}
    \mathcal{C}_{S,S'}
    \mathcal{F}_{S,S'}^{\mu,T} 
    \left[
    \tau_{S} - \frac{v_{S'}}{v_S}\tau_{S'}\cos\theta_{\vex{v}}
    \right]
    &= 1.
\end{align}
Above, $\rho_M$ is the mass density of the system, $D$ is the deformation potential, and $v_p$ is the phonon velocity. The function $\tilde{\Delta}_{S,S'}$ enforces the conservation of energy and momentum and defines the scattering manifold, while $\mathcal{C}_{S,S'}$ are ``matrix elements" encoding geometric wavefunction overlap data \cite{Hwang2008,Wu2019_phonon,Davis_2022,Davis_2023}. The dependencies on temperature and chemical potential are contained in $\mathcal{F}_{S,S'}^{\mu,T}$, which encodes the thermal occupation data for the fermion and phonon states. These are given explicitly in the SM \cite{SM}. On a finite momentum grid, Eq.~(\ref{RelaxationLengthSelfConsistency}) is a matrix inversion problem that can be solved for the relaxation times, $\tau_S$. Equation~(\ref{ResistivityDefinition}) then gives the resistivity.

%%%%%%%%%%%%%%%%%%%%%%%%%%%%%%%%%%%%%%%%%%%%%%%%%%%%%%%%%
%%%%%%%%%%%%%%%%%%%%%%%%%%%%%%%%%%%%%%%%%%%%%%%%%%%%%%%%%
%%%%%%%%%%%%%%%%%%%%%%%%%%%%%%%%%%%%%%%%%%%%%%%%%%%%%%%%%
%%%%%%%%%%%%%%%%%%%%%%%%%%%%%%%%%%%%%%%%%%%%%%%%%%%%%%%%%
%%%%%%%%%%%%%%%%%%%%%%%%%%%%%%%%%%%%%%%%%%%%%%%%%%%%%%%%%
%%%%%%%%%%%%%%%%%%%%%%%%%%%%%%%%%%%%%%%%%%%%%%%%%%%%%%%%%
%%%%%%%%%%%%%%%%%%%%%%%%%%%%%%%%%%%%%%%%%%%%%%%%%%%%%%%%%

% Equipartition regime

%%%%%%%%%%%%%%%%%%%%%%%%%%%%%%%%%%%%%%%%%%%%%%%%%%%%%%%%%
%%%%%%%%%%%%%%%%%%%%%%%%%%%%%%%%%%%%%%%%%%%%%%%%%%%%%%%%%
%%%%%%%%%%%%%%%%%%%%%%%%%%%%%%%%%%%%%%%%%%%%%%%%%%%%%%%%%
%%%%%%%%%%%%%%%%%%%%%%%%%%%%%%%%%%%%%%%%%%%%%%%%%%%%%%%%%
%%%%%%%%%%%%%%%%%%%%%%%%%%%%%%%%%%%%%%%%%%%%%%%%%%%%%%%%%
%%%%%%%%%%%%%%%%%%%%%%%%%%%%%%%%%%%%%%%%%%%%%%%%%%%%%%%%%
%%%%%%%%%%%%%%%%%%%%%%%%%%%%%%%%%%%%%%%%%%%%%%%%%%%%%%%%%
\begin{figure}[t!]
\includegraphics[angle=0,width=.47\textwidth]{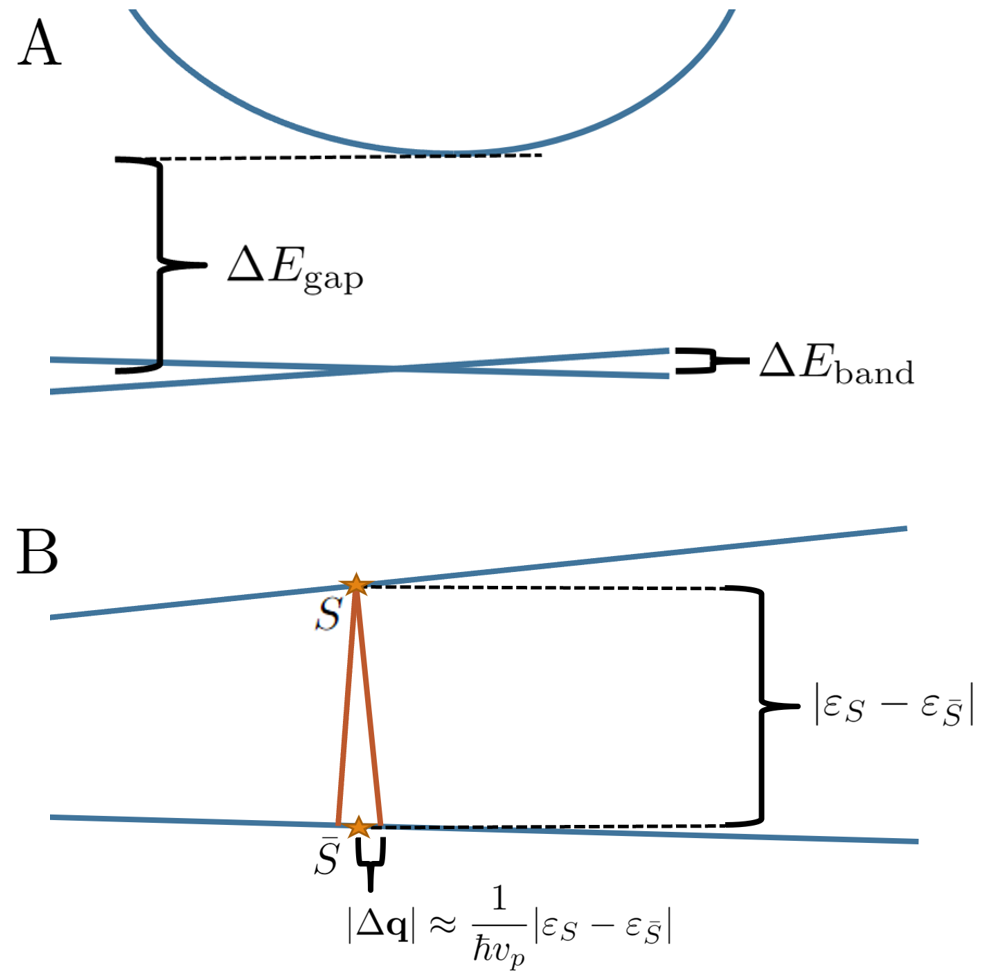}
\caption{In (A) we depict the scenario of an NFB system isolated from the rest of the band structure by a sizable energy gap. Phonon scattering can produce a $\rho \propto T^2$ resistivity power law for $\Delta E_{\text{band}} \ll k_B T \ll \Delta E_{\text{gap}}$ in this band geometry. In (B), we depict the band geometry of the ultrasubsonic limit, defined by $v_F \ll v_p$ and small band separation.}
\label{SubsonicSchematic2}
\end{figure}

\textit{Equipartition regime: $\rho \propto T^2$}.---When $T \gg T_{BG}$ (equipartition regime), the inverse relaxation times scale linearly with temperature:
\begin{align}
\label{EquipartitionScaling}
    \tau_S = \frac{c_S}{k_BT} + \mathcal{O}\left(\frac{\Delta\e}{(k_BT)^3}\right),
\end{align}
for some $(T,\mu)$-independent proportionality constants $c_S$ that depend on the details of the system. In the case of NFBs separated from all other bands by a large energy gap, there is a mid-$T$ regime in which $T$ is larger than the NFB bandwidth, yet $T$ is still small compared to the gap between the NFBs and the other bands [Fig.~\ref{SubsonicSchematic2} (A)]. In this case, we may neglect bands other than the NFBs, apply the equipartition scaling for the relaxation times [Eq.~(\ref{EquipartitionScaling})], and expand the thermal weighting functions ($\cosh$) in the formula for the resistivity [Eq.~(\ref{ResistivityDefinition})], giving
\begin{align}
\label{EquipartitionT2}
    \frac{\delta^{ij}}{\rho} 
    &= 
    \frac{1}{4}
    \frac{e^2}{(k_BT)^2}\frac{1}{\mathcal{A}}
    \sum_{S}
    c_{S}
    v^i_{S}v^j_{S}
    +
    \mathcal{O}\left(\frac{1}{(k_BT)^4}\right).
\end{align}
We thus find a mid-$T$, $\rho \propto T^2$ scaling regime due entirely to phonon scattering. This is noteworthy since $\rho \propto T^2$ scaling is generally seen as the hallmark of transport dominated by electron-electron scattering in a Fermi liquid \cite{AshcroftAndMermin, Ziman, Coleman2015introduction}, but the ``mid-$T$" regime generated by NFB phonon scattering has exactly the same $T^2$ dependence.

%%%%%%%%%%%%%%%%%%%%%%%%%%%%%%%%%%%%%%%%%%%%%%%%%%%%%%%%%
%%%%%%%%%%%%%%%%%%%%%%%%%%%%%%%%%%%%%%%%%%%%%%%%%%%%%%%%%
%%%%%%%%%%%%%%%%%%%%%%%%%%%%%%%%%%%%%%%%%%%%%%%%%%%%%%%%%
%%%%%%%%%%%%%%%%%%%%%%%%%%%%%%%%%%%%%%%%%%%%%%%%%%%%%%%%%
%%%%%%%%%%%%%%%%%%%%%%%%%%%%%%%%%%%%%%%%%%%%%%%%%%%%%%%%%
%%%%%%%%%%%%%%%%%%%%%%%%%%%%%%%%%%%%%%%%%%%%%%%%%%%%%%%%%
%%%%%%%%%%%%%%%%%%%%%%%%%%%%%%%%%%%%%%%%%%%%%%%%%%%%%%%%%

% Ultrasubsonic kinetic theory

%%%%%%%%%%%%%%%%%%%%%%%%%%%%%%%%%%%%%%%%%%%%%%%%%%%%%%%%%
%%%%%%%%%%%%%%%%%%%%%%%%%%%%%%%%%%%%%%%%%%%%%%%%%%%%%%%%%
%%%%%%%%%%%%%%%%%%%%%%%%%%%%%%%%%%%%%%%%%%%%%%%%%%%%%%%%%
%%%%%%%%%%%%%%%%%%%%%%%%%%%%%%%%%%%%%%%%%%%%%%%%%%%%%%%%%
%%%%%%%%%%%%%%%%%%%%%%%%%%%%%%%%%%%%%%%%%%%%%%%%%%%%%%%%%
%%%%%%%%%%%%%%%%%%%%%%%%%%%%%%%%%%%%%%%%%%%%%%%%%%%%%%%%%
%%%%%%%%%%%%%%%%%%%%%%%%%%%%%%%%%%%%%%%%%%%%%%%%%%%%%%%%%

\textit{Ultrasubsonic kinetic theory}.---We now consider electron-phonon scattering in a limit that allows transparent analytical results. We consider a two-band model in the double-limit of small $v_F/v_p$ and small band separation, which we call the ``ultrasubsonic (USS) limit" [see Fig.~\ref{SubsonicSchematic2} (B)]. For a given Bloch state ($S$, with energy $\e$), its scattering manifold is a small loop on the opposite band, surrounding its ``compliment", the point on the opposite band with the same momentum as the original state ($\bar{S}$, with energy $\bar{\e}$). In the USS limit, the scattering manifold is well-approximated by a circle of radius $|\e-\bar{\e}|/(\hbar v_p)$, centered at $\bar{S}$. The energy variation along the scattering manifold is then roughly $(v_F/v_p)|\e-\bar{\e}|,$ which is double-suppressed in the USS limit. We thus make the approximations $\mathcal{F}^{\mu,T}(S,S') \approx \mathcal{F}^{\mu,T}(S,\bar{S})$ and $\tau(S') \approx \tau(\bar{S})$ for $S'$ along the scattering manifold, approximating the relaxation times and the thermal occupancy function along the scattering manifold by their values at the compliment point. 

Using the approximations discussed above and taking the thermodynamic limit (continuum limit in momentum space) of Eq.~(\ref{RelaxationLengthSelfConsistency}), we have
\begin{align}
    \label{DerivationRootEquation}
    1
    &=
    \frac{D^2}{2 \rho_M \hbar^3 v_p^4}
    |\e-\bar{\e}| \mathcal{F}^{\mu,T}_{S,\bar{S}}
    \left[
    X_S \tau_{S} 
    - 
    \tilde{X}_S \tau_{\bar{S}}
    \right],
\end{align}
\begin{widetext}
\noindent
where the $(T,\mu)$-independent factors $X_{S}$ and $\tilde{X}_S$ encode wavefunction overlap and band geometry data. They are given in the SM \cite{SM}. We may combine Eq.~(\ref{DerivationRootEquation}) for the state $S$ with its analogue for the compliment state $\bar{S}$, into a 2x2 matrix equation determining both $\tau(S)$ and $\tau(\bar{S})$, which we solve analytically for the USS relaxation times:
\begin{align}
    \tau_S 
    &=
\label{GeneralSolutionTau}    
    \frac{2 \rho_M \hbar^3 v_p^4}{D^2} 
    \frac{2}{|\e-\bar{\e}|^2}
    \sinh\left(\frac{|\e-\bar{\e}|}{2 k_B T}\right)
    \frac{1}{X_S X_{\bar{S}} - \tilde{X}_S \tilde{X}_{\bar{S}}}
    \left[
    X_{\bar{S}}
    \frac{\cosh\left(
    \frac{\bar{\e}-\mu}{2k_BT}
    \right)}
    {\cosh\left(
    \frac{\e-\mu}{2k_BT}
    \right)}
    +
    \tilde{X}_S
    \frac{\cosh\left(
    \frac{\e-\mu}{2k_BT}
    \right)}
    {\cosh\left(
    \frac{\bar{\e}-\mu}{2k_BT}
    \right)}
    \right].
\end{align}
Combining the relaxation time formula [Eq.~(\ref{GeneralSolutionTau})] with the resistivity formula [Eq.~(\ref{ResistivityDefinition})] gives the master formula for the resistivity of the ultra-subsonic fermion system. We present this in the SM \cite{SM}. In the case of a particle-hole (PH) symmetric system, the formulae for the USS resistivity simplifies: 
\begin{align}
\label{PHSymmetricResistivity}
    \frac{\delta^{ij}}{\rho} 
    &= 
    \frac{1}{4}
    \frac{e^2}{k_BT}
    \frac{\rho_M \hbar^3 v_p^4}{D^2}
    \left\lgroup
    \begin{aligned}
    &
    \frac{1}{\mathcal{A}}
    \sum_{S}
    \frac{v^i_{S}v^j_{S}}{X_S - \tilde{X}_S}
    \frac{1}{\e^2}
    \frac{\sinh\left(\frac{|\e|}{k_B T}\right)}
    {\cosh\left(\frac{\e+\mu}{2k_B T}\right)\cosh\left(\frac{\e-\mu}{2k_B T}\right)}
    \\
    +
    &
    \frac{1}{2}
    \sinh\left(\frac{|\mu|}{k_B T}\right)^2
    \frac{1}{\mathcal{A}}
    \sum_{S}
    \frac{v^i_{S}v^j_{S} X_S}{X^2_S - \tilde{X}^2_S}
    \frac{1}{\e^2}
    \left[
    \frac{
    \sinh\left(\frac{|\e|}{k_B T}\right)
    }
    {
    \cosh\left(\frac{\e+\mu}{2k_B T}\right)
    \cosh\left(\frac{\e-\mu}{2k_B T}\right)
    }
    \right]^3
    \end{aligned}
    \right\rgroup.
\end{align}
\end{widetext}

\begin{figure}[b!]
\includegraphics[angle=0,width=.47\textwidth]{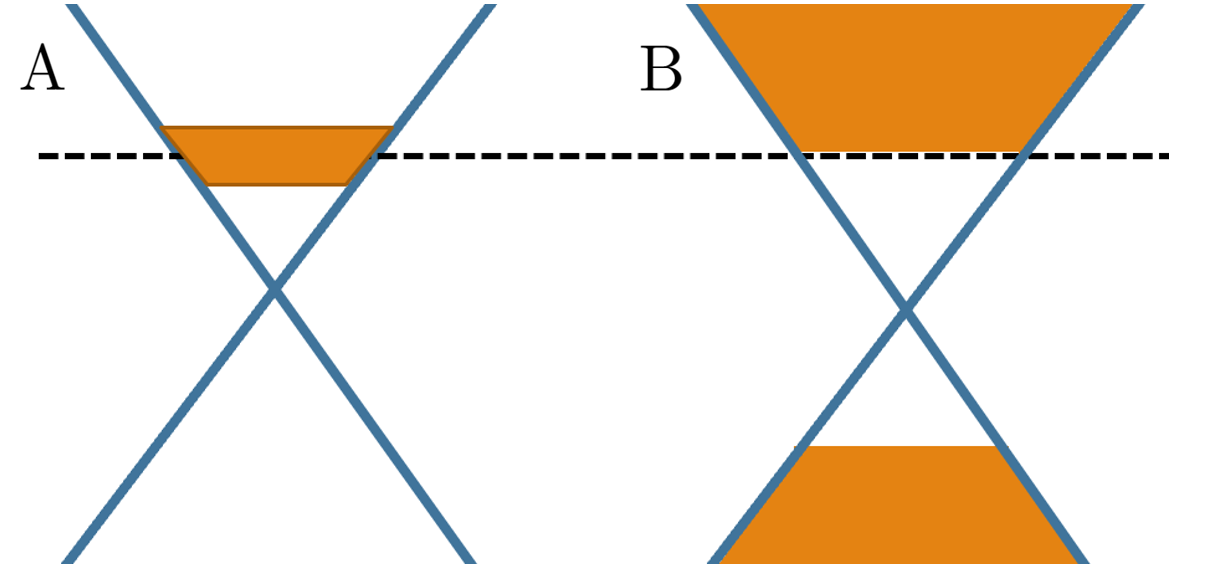}
\caption{We depict the difference in the physics of low-$T$ transport in subsonic and supersonic fermion systems. In each figure, the band structure of the fermions is depicted in blue, and the orange filling indicates the energies of the electrons that contribute meaningfully to transport. The dotted line denotes the Fermi level. In (A), we show the scenario for supersonic fermions, in which only the states in the immediate vicinity of the Fermi surface contribute to transport. In (B), we illustrate the different physics of subsonic fermion scattering, where the states contributing meaningfully to low-$T$ transport are those satisfying $|\e| > |\mu|$. In this picture, states very far from the Fermi level can contribute much more than states even just barely below it.}
\label{SubsonicSchematic3}
\end{figure}

In the limit $k_B T \gg |\e-\bar{\e}|$, expanding the $\sinh$ factor in Eq.~(\ref{GeneralSolutionTau}) gives the expected $\tau \propto T^{-1}$ scattering rate of the equipartition regime. However, when $k_B T \ll |\e-\bar{\e}|$, instead of the usual crossover to a $\tau \propto T^{-4}$ power law, we find an exponential blow up of the relaxation times, in line with the physical picture in which the entire scattering manifold becomes thermally inaccessible.

%%%%%%%%%%%%%%%%%%%%%%%%%%%%%%%%%%%%%%%%%%%%%%%%%%%%%%%%%
%%%%%%%%%%%%%%%%%%%%%%%%%%%%%%%%%%%%%%%%%%%%%%%%%%%%%%%%%
%%%%%%%%%%%%%%%%%%%%%%%%%%%%%%%%%%%%%%%%%%%%%%%%%%%%%%%%%
%%%%%%%%%%%%%%%%%%%%%%%%%%%%%%%%%%%%%%%%%%%%%%%%%%%%%%%%%
%%%%%%%%%%%%%%%%%%%%%%%%%%%%%%%%%%%%%%%%%%%%%%%%%%%%%%%%%
%%%%%%%%%%%%%%%%%%%%%%%%%%%%%%%%%%%%%%%%%%%%%%%%%%%%%%%%%
%%%%%%%%%%%%%%%%%%%%%%%%%%%%%%%%%%%%%%%%%%%%%%%%%%%%%%%%%

% Extreme low-$T$ limit

%%%%%%%%%%%%%%%%%%%%%%%%%%%%%%%%%%%%%%%%%%%%%%%%%%%%%%%%%
%%%%%%%%%%%%%%%%%%%%%%%%%%%%%%%%%%%%%%%%%%%%%%%%%%%%%%%%%
%%%%%%%%%%%%%%%%%%%%%%%%%%%%%%%%%%%%%%%%%%%%%%%%%%%%%%%%%
%%%%%%%%%%%%%%%%%%%%%%%%%%%%%%%%%%%%%%%%%%%%%%%%%%%%%%%%%
%%%%%%%%%%%%%%%%%%%%%%%%%%%%%%%%%%%%%%%%%%%%%%%%%%%%%%%%%
%%%%%%%%%%%%%%%%%%%%%%%%%%%%%%%%%%%%%%%%%%%%%%%%%%%%%%%%%
%%%%%%%%%%%%%%%%%%%%%%%%%%%%%%%%%%%%%%%%%%%%%%%%%%%%%%%%%

\textit{Extreme low-$T$ limit}.---In the PH symmetric case [Eq.~\ref{PHSymmetricResistivity}], the only $T$-dependent factor in the Brillouin zone summations is is the kernel
\begin{align}
\label{ThermalKernel}    
    K(\e,\mu,T) 
    \equiv
    \frac{
    \sinh\left(\frac{|\e|}{k_B T}\right)
    }
    {
    \cosh\left(\frac{\e+\mu}{2k_B T}\right)
    \cosh\left(\frac{\e-\mu}{2k_B T}\right)
    }.
\end{align}
This factor represents the competition at low-$T$ between the divergence of the relaxation times and the thermodynamic suppression of states far from the Fermi level. In the extreme low-$T$ limit, where all the $\sinh$ and $\cosh$ functions blow up exponentially, we simply have
\begin{align}
\label{LowTLimitReplacement}
    K(\e, \mu, T \rightarrow 0)
    \rightarrow
    2\Theta[|\e| - |\mu|].
\end{align}
The thermal kernel $K$ determines which Bloch states meaningfully contribute to transport. Equation~(\ref{LowTLimitReplacement}) suggests that all states with $|\mu| < |\e|$ contribute equally, while states with $|\mu| > |\e|$ do not contribute, even though they may be close to the Fermi energy. We emphasize that this is radically different than the usual paradigm, where low-$T$ transport is almost entirely determined by states in the immediate vicinity of the Fermi level. This is depicted in Fig.~\ref{SubsonicSchematic3}.

Applying Eq.~(\ref{LowTLimitReplacement}), the two summations in Eq.~(\ref{PHSymmetricResistivity}) are simply $T$-independent constants ($\mathcal{C}_1$, $\mathcal{C}_2$), and the extreme low-$T$ expression for the resistivity is
\begin{align}
\label{LowTPHResistivity2}
    \rho
    &\approx
    \frac{\hbar D^2}{4e^2 \rho_M (\hbar v_p)^4}
    \frac{k_BT}{\mathcal{C}_1 + \mathcal{C}_2\sinh[|\mu|/(k_B T)]^2}.
\end{align}
From Eq.~(\ref{LowTPHResistivity2}), it is apparent that at the charge neutrality point ($\mu = 0$), we have purely linear-in-$T$ phonon-induced resistivity down to $T = 0$. On the other hand, when $\mu \neq 0$, then the low-$T$ resistivity is proportional to the factor $\rho \propto \exp[-2|\mu|/(k_B T)]$, and is exponentially suppressed when $k_B T \ll 2|\mu|$, giving a crossover to perfect conductivity. Comparing this with the physics of supersonic fermions, we find
\begin{align}
    k_BT_{\text{crossover}} = 2|\mu| = \frac{v_F}{v_p} k_B T^{\text{trad}}_{BG},
\end{align}
where $T^{\text{trad}}_{BG} \equiv 2 v_p k_F$ gives the traditional lower bound for the regime of linear-in-$T$ resistivity based on the usual Bloch-Gr\"{u}neisen paradigm. The crossover temperature is parametrically suppressed by the small parameter, $v_F/v_p$. The subsonic case is thus expected to host linear-in-$T$ resistivity scaling down to a significantly lower temperature than one would estimate based on intuition from supersonic fermion scattering.

%%%%%%%%%%%%%%%%%%%%%%%%%%%%%%%%%%%%%%%%%%%%%%%%%%%%%%%%%
%%%%%%%%%%%%%%%%%%%%%%%%%%%%%%%%%%%%%%%%%%%%%%%%%%%%%%%%%
%%%%%%%%%%%%%%%%%%%%%%%%%%%%%%%%%%%%%%%%%%%%%%%%%%%%%%%%%
%%%%%%%%%%%%%%%%%%%%%%%%%%%%%%%%%%%%%%%%%%%%%%%%%%%%%%%%%
%%%%%%%%%%%%%%%%%%%%%%%%%%%%%%%%%%%%%%%%%%%%%%%%%%%%%%%%%
%%%%%%%%%%%%%%%%%%%%%%%%%%%%%%%%%%%%%%%%%%%%%%%%%%%%%%%%%
%%%%%%%%%%%%%%%%%%%%%%%%%%%%%%%%%%%%%%%%%%%%%%%%%%%%%%%%%

% Conclusion

%%%%%%%%%%%%%%%%%%%%%%%%%%%%%%%%%%%%%%%%%%%%%%%%%%%%%%%%%
%%%%%%%%%%%%%%%%%%%%%%%%%%%%%%%%%%%%%%%%%%%%%%%%%%%%%%%%%
%%%%%%%%%%%%%%%%%%%%%%%%%%%%%%%%%%%%%%%%%%%%%%%%%%%%%%%%%
%%%%%%%%%%%%%%%%%%%%%%%%%%%%%%%%%%%%%%%%%%%%%%%%%%%%%%%%%
%%%%%%%%%%%%%%%%%%%%%%%%%%%%%%%%%%%%%%%%%%%%%%%%%%%%%%%%%
%%%%%%%%%%%%%%%%%%%%%%%%%%%%%%%%%%%%%%%%%%%%%%%%%%%%%%%%%
%%%%%%%%%%%%%%%%%%%%%%%%%%%%%%%%%%%%%%%%%%%%%%%%%%%%%%%%%

\textit{Concluding discussion}.---Phonon scattering processes in subsonic fermion systems are necessarily interband transitions, implying the existence of a nonzero, minimum energy transfer allowed by kinematics. At temperatures below this energy scale, phonon scattering is suppressed and Bloch states become long-lived, with important effects on the low-$T$ transport physics. Applying standard Boltzmann kinetic theory of acoustic phonon scattering to nearly flat band (NFB) systems, we find that this underlies a robust, linear-in-$T$ scaling of the resistivity down to temperatures far lower than the Bloch-Gr\"{u}neisen paradigm of supersonic fermion scattering would suggest. This result provides a concrete mechanism for the linear-in-$T$ resistivity over a wide range of temperatures based only on familiar concepts of solid state physics and the distinct kinematics of NFB systems. In particular, it provides an alternative theoretical explanation for low-$T$ ``strange metal" resistivity scaling in NFB systems. Our theory also predicts regimes of perfect conductivity and $\rho \propto T^2$ scaling which compete with the $\rho \propto T$ regime, which are not commonly associated with phonon physics. Both $\rho \propto T^2$ and low-$T$ $\rho \propto T$ behaviors are reported in MATBLG \cite{Jaoui_2022}. Thus, low-$T$ $\rho \propto T$, $\rho \propto T^2$, and perfect conductivity regimes may all arise in NFB systems from the same universal phonon-scattering physics, providing a possible explanation for these reported observations in MATBLG.

Our theory of USS fermion transport is general and could apply widely to NFB systems in other 2D heterostructures or heavy fermion materials. However, our results are particularly germane to the ongoing debate on the presence of a strange metal state in MATBLG. Using the Bistritzer-MacDonald (BM) model for the non-interacting band structure of MATBLG, we see that much of the band is made up of subsonic states ($\approx 92\%$). Large sections of the NFBs have small band separation ($\leq 0.0005\ eV$) and small $v_F/v_p$ ($\leq 0.02$), rendering USS theory highly applicable. We discuss this more quantitatively in the SM \cite{SM}. We emphasize that a direct quantitative application of USS theory to the BM model could be of limited utility since the bands are expected to be renormalized by interaction effects, strain, and twist disorder.

%%%%%%%%%%%%%%%%%%%%%%%%%%%%%%%%%%%%%%%%%%%%%%%%%%%%%%%%%
%%%%%%%%%%%%%%%%%%%%%%%%%%%%%%%%%%%%%%%%%%%%%%%%%%%%%%%%%
%%%%%%%%%%%%%%%%%%%%%%%%%%%%%%%%%%%%%%%%%%%%%%%%%%%%%%%%%
%%%%%%%%%%%%%%%%%%%%%%%%%%%%%%%%%%%%%%%%%%%%%%%%%%%%%%%%%
%%%%%%%%%%%%%%%%%%%%%%%%%%%%%%%%%%%%%%%%%%%%%%%%%%%%%%%%%
%%%%%%%%%%%%%%%%%%%%%%%%%%%%%%%%%%%%%%%%%%%%%%%%%%%%%%%%%
%%%%%%%%%%%%%%%%%%%%%%%%%%%%%%%%%%%%%%%%%%%%%%%%%%%%%%%%%

% Acknowledgments 

%%%%%%%%%%%%%%%%%%%%%%%%%%%%%%%%%%%%%%%%%%%%%%%%%%%%%%%%%
%%%%%%%%%%%%%%%%%%%%%%%%%%%%%%%%%%%%%%%%%%%%%%%%%%%%%%%%%
%%%%%%%%%%%%%%%%%%%%%%%%%%%%%%%%%%%%%%%%%%%%%%%%%%%%%%%%%
%%%%%%%%%%%%%%%%%%%%%%%%%%%%%%%%%%%%%%%%%%%%%%%%%%%%%%%%%
%%%%%%%%%%%%%%%%%%%%%%%%%%%%%%%%%%%%%%%%%%%%%%%%%%%%%%%%%
%%%%%%%%%%%%%%%%%%%%%%%%%%%%%%%%%%%%%%%%%%%%%%%%%%%%%%%%%
%%%%%%%%%%%%%%%%%%%%%%%%%%%%%%%%%%%%%%%%%%%%%%%%%%%%%%%%%

\acknowledgments

We thank Fengcheng Wu for helpful discussions. This work is supported by the Laboratory for Physical Sciences.

\bibliography{BIB}
\end{document}

\title{The kinetic theory of ultra-subsonic fermion systems and applications to magic angle twisted bilayer graphene\\
SUPPLEMENTAL MATERIAL\\}
\author{Seth M.\ Davis}
\email{smdavis1@umd.edu}
\affiliation{Condensed Matter Theory Center and Joint Quantum Institute, Department of Physics, University of Maryland, College Park, MD 20742, USA}
\author{Sankar\ Das\ Sarma}
\affiliation{Condensed Matter Theory Center and Joint Quantum Institute, Department of Physics, University of Maryland, College Park, MD 20742, USA}
\date{\today}
\maketitle

%%%%%%%%%%%%%%%%%%%%%%%%%%%%%%%%%%%%%%%%%%%%%%%%%%%%%%%%%%%%%%%%%%%%%%%%%%%%%%%%
%%%%%%%%%%%%%%%%%%%%%%%%%%%%%%%%%%%%%%%%%%%%%%%%%%%%%%%%%%%%%%%%%%%%%%%%%%%%%%%%
%%%%%%%%%%%%%%%%%%%%%%%%%%%%%%%%%%%%%%%%%%%%%%%%%%%%%%%%%%%%%%%%%%%%%%%%%%%%%%%%
%%%%%%%%%%%%%%%%%%%%%%%%%%%%%%%%%%%%%%%%%%%%%%%%%%%%%%%%%%%%%%%%%%%%%%%%%%%%%%%%
%%%%%%%%%%%%%%%%%%%%%%%%%%%%%%%%%%%%%%%%%%%%%%%%%%%%%%%%%%%%%%%%%%%%%%%%%%%%%%%%
%%%%%%%%%%%%%%%%%%%%%%%%%%%%%%%%%%%%%%%%%%%%%%%%%%%%%%%%%%%%%%%%%%%%%%%%%%%%%%%%
%%%%%%%%%%%%%%%		Supplementary Material		%%%%%%%%%%%%%%%%%%%%%%%%
%%%%%%%%%%%%%%%		Supplementary Material		%%%%%%%%%%%%%%%%%%%%%%%%
%%%%%%%%%%%%%%%		Supplementary Material		%%%%%%%%%%%%%%%%%%%%%%%%
%%%%%%%%%%%%%%%%%%%%%%%%%%%%%%%%%%%%%%%%%%%%%%%%%%%%%%%%%%%%%%%%%%%%%%%%%%%%%%%%
%%%%%%%%%%%%%%%%%%%%%%%%%%%%%%%%%%%%%%%%%%%%%%%%%%%%%%%%%%%%%%%%%%%%%%%%%%%%%%%%
%%%%%%%%%%%%%%%%%%%%%%%%%%%%%%%%%%%%%%%%%%%%%%%%%%%%%%%%%%%%%%%%%%%%%%%%%%%%%%%%
%%%%%%%%%%%%%%%%%%%%%%%%%%%%%%%%%%%%%%%%%%%%%%%%%%%%%%%%%%%%%%%%%%%%%%%%%%%%%%%%
%%%%%%%%%%%%%%%%%%%%%%%%%%%%%%%%%%%%%%%%%%%%%%%%%%%%%%%%%%%%%%%%%%%%%%%%%%%%%%%%
%%%%%%%%%%%%%%%%%%%%%%%%%%%%%%%%%%%%%%%%%%%%%%%%%%%%%%%%%%%%%%%%%%%%%%%%%%%%%%%%
%%%%%%%%%%%%%%%%%%%%%%%%%%%%%%%%%%%%%%%%%%%%%%%%%%%%%%%%%%%%%%%%%%%%%%%%%%%%%%%%

\newpage \clearpage

\onecolumngrid

%\begin{center}
%	{\large
%	Fractionalization Waves in Two-dimensional Dirac Fermions: \\
%	Quantum Imprint from One Dimension
%	\vspace{4pt}
%	\\
%	SUPPLEMENTAL MATERIAL
%	}
%\end{center}
%%%%%%%%%% Prefix a "S" to all equations, figures, tables and reset the counter %%%%%%%%%%
\counterwithin{figure}{section}
\counterwithin{equation}{section}
\makeatletter
\setcounter{equation}{0}
\setcounter{figure}{0}
\setcounter{table}{0}
\setcounter{page}{1}
\renewcommand{\theequation}{S\arabic{equation}}
\renewcommand{\thefigure}{S\arabic{figure}}
\renewcommand{\bibnumfmt}[1]{[S#1]}
\renewcommand{\citenumfont}[1]{S#1}
%%%%%%%%%% Prefix a "S" to all equations, figures, tables 

\begingroup
\hypersetup{linkbordercolor=white}
\tableofcontents
\endgroup

%%%%%%%%%%%%%%%%%%%%%%%%%%%%%%%%%%%%%%%%%%%%%%%%%%%%%%%%%
%%%%%%%%%%%%%%%%%%%%%%%%%%%%%%%%%%%%%%%%%%%%%%%%%%%%%%%%%
%%%%%%%%%%%%%%%%%%%%%%%%%%%%%%%%%%%%%%%%%%%%%%%%%%%%%%%%%
%%%%%%%%%%%%%%%%%%%%%%%%%%%%%%%%%%%%%%%%%%%%%%%%%%%%%%%%%
%%%%%%%%%%%%%%%%%%%%%%%%%%%%%%%%%%%%%%%%%%%%%%%%%%%%%%%%%
%%%%%%%%%%%%%%%%%%%%%%%%%%%%%%%%%%%%%%%%%%%%%%%%%%%%%%%%%
%%%%%%%%%%%%%%%%%%%%%%%%%%%%%%%%%%%%%%%%%%%%%%%%%%%%%%%%%

%%%%%%%%%%%%%%%%%%%%%%%%%%%%%%%%%%%%%%%%%%%%%%%%%%%%%%%%%
%%%%%%%%%%%%%%%%%%%%%%%%%%%%%%%%%%%%%%%%%%%%%%%%%%%%%%%%%
%%%%%%%%%%%%%%%%%%%%%%%%%%%%%%%%%%%%%%%%%%%%%%%%%%%%%%%%%
% Microscopic model
%%%%%%%%%%%%%%%%%%%%%%%%%%%%%%%%%%%%%%%%%%%%%%%%%%%%%%%%%
%%%%%%%%%%%%%%%%%%%%%%%%%%%%%%%%%%%%%%%%%%%%%%%%%%%%%%%%%
%%%%%%%%%%%%%%%%%%%%%%%%%%%%%%%%%%%%%%%%%%%%%%%%%%%%%%%%%

\section{I.\ Microscopic model}
\label{SM-MicroscopicModel}
We consider a general, subsonic two-band fermion system coupled via the standard deformation potential \cite{Ziman, SankarGrapheneKT1, Coleman2015introduction} to longitudinal acoustic phonons in the Debye approximation. Thus, our model is described by the Hamiltonian
\begin{align}
\label{TotalHamiltonian}
    H &= H^{e} + H^{ph} + H^{e-ph},
    \\
\label{PhononHamiltonian}
    H^{ph} &= \sum_{l,\vex{q}}\hbar\omega_{\vex{q}}a_{l,\vex{q}}^\dagger a_{l,\vex{q}},
    \\
\label{CouplingHamiltonian}
    H^{e-ph} &= 
    \sqrt{\frac{D^2 \hbar}{2\rho_M \mathcal{A}}}
    \sum_{l,\vex{q}}
    \frac{\hat{n}_{\vex{q},l}}{\sqrt{\omega_{\vex{q}}}}(-i\vex{q}\cdot\hat{e}_{\vex{q}})(a_{\vex{q},l} + a_{-\vex{q},l}^\dagger).
\end{align}
Above, $\mathcal{A}$ is the area of the system, $\rho_M$ is the mass density of the system, $D$ is the deformation potential, $\hat{e}_{\vex{q}}$ is the displacement unit vector of the phonon, and $\omega_{\vex{q}}$ is the phonon dispersion. Finally, the electron density operator is 
\begin{align}
\label{DensityOperator}
    \hat{n}_{\vex{q},l} \equiv \sum_{\vex{k}}c^{\dagger}_{(\vex{k}+\vex{q}),l}c_{\vex{k},l},
\end{align}
where $c_{\vex{k}}^\dagger \equiv c^\dagger_{s,\xi,\sigma,l,\vex{k}}$ creates an electron with momentum $\vex{k}$, spin $s$, valley $\xi$, sublattice $\sigma$, and layer $l$. Summations over the unwritten indices $s,\xi,$ and $\sigma$ are implicit in Eq.~(\ref{DensityOperator}). We use the Debye approximation $\omega_{\vex{q}} \approx v_{p}|\vex{q}|$, where $v_p$ is the phonon (or sound) velocity of graphene. (Since our focus will be on the extreme low-$T$ limit, the Debye approximation is particularly justified.) The standard values of the above parameters for monolayer graphene are $D = 25$ eV, $\rho_M = 7.6 \cdot 10^{-8} g/cm^2$, and $v_p = 2.0 \cdot 10^6 cm/s$, \cite{SankarGrapheneKT1, SankarGrapheneKT2, Wu2019_phonon, Efetov_2010}. Using these values in the theory gives accurate quantitative predictions for transport in many layered graphene systems \cite{Wu2019_phonon,Davis_2022,Davis_2023}.

The single-particle electron Hamiltonian $H^e$ determines the electronic band structure, which we will not specify in order to preserve generality. However, we will consider a two-band system and assume both that $v_F \ll v_p$ and that the separation between the bands is small. We will call the limit in which both these assumptions are valid the ``ultrasubsonic" (USS) limit.

%%%%%%%%%%%%%%%%%%%%%%%%%%%%%%%%%%%%%%%%%%%%%%%%%%%%%%%%%
%%%%%%%%%%%%%%%%%%%%%%%%%%%%%%%%%%%%%%%%%%%%%%%%%%%%%%%%%
%%%%%%%%%%%%%%%%%%%%%%%%%%%%%%%%%%%%%%%%%%%%%%%%%%%%%%%%%
% Boltzmann kinetic theory details
%%%%%%%%%%%%%%%%%%%%%%%%%%%%%%%%%%%%%%%%%%%%%%%%%%%%%%%%%
%%%%%%%%%%%%%%%%%%%%%%%%%%%%%%%%%%%%%%%%%%%%%%%%%%%%%%%%%
%%%%%%%%%%%%%%%%%%%%%%%%%%%%%%%%%%%%%%%%%%%%%%%%%%%%%%%%%

\section{II.\ Boltzmann kinetic theory details}
\label{SM-BoltzmannKineticTheoryDetails}
We provide additional details on the Boltzmann kinetic theory framework for longitudinal acoustic phonon scattering. As stated in the main text [Eqs.~(1-2)], the resistivity of the system follows from the relaxation times via
\begin{align}
\label{ResistivityDefinition}
    \frac{\delta^{ij}}{\rho}
    &=
    \frac{1}{4}
    \frac{e^2}{k_B T}
    \frac{1}{\mathcal{A}}
    \sum_{S}
    \frac{
    \tau_{S}
    v^i_{S}v^j_{S}
    }
    {
    \cosh\left(
    \frac{\e-\mu}{2 k_B T}
    \right)^2
    },
\end{align}
and the relaxation times are determined from the self-consistent integral equation
\begin{align}
    \label{RelaxationLengthSelfConsistency}
    \frac{\pi D^2}{\hbar \rho_M v_p^2}
    \frac{1}{\mathcal{A}}
    \sum_{S'}
    \tilde{\Delta}_{S,S'}
    \mathcal{C}_{S,S'}
    \mathcal{F}_{S,S'}^{\mu,T} 
    \left[
    \tau_{S} - \frac{v_{S'}}{v_S}\tau_{S'}\cos\theta_{\vex{v}}
    \right]
    &= 1.
\end{align}
In a layered graphene system, we note that the summation over states $S$ in Eq.~(\ref{ResistivityDefinition}) includes a 4-fold degeneracy over spin and valley degrees of freedom.
In Eq.~(\ref{RelaxationLengthSelfConsistency}),
\begin{align}
\label{SMCondition}
     \tilde{\Delta}_{S,S'} \equiv \delta(|\e'-\e|-\hbar v_p |\vex{k}'-\vex{k}|)
\end{align}
enforces the conservation laws and defines the scattering manifold. Note that the energy-momentum conservation assumes the Debye dispersion for the phonons. The function
\begin{align}
    \mathcal{F}_{S,S'}^{\mu,T}
    &\equiv 
    \frac{|\e'-\e|}{2}
    \cfrac{1}{\sinh\left[\cfrac{|\e'-\e|}{2 k_B T}\right]}
    \cfrac{\cosh\left[\cfrac{\e-\mu}{2 k_B T}\right]}
    {\cosh\left[\cfrac{\e'-\mu}{2 k_B T}\right]}.
\end{align}
encodes all the thermodynamics, including the occupation data for the fermions and phonons and all dependence on $\mu$ and $T$. Finally, we have the geometric wavefunction matrix element
\begin{align}
\label{OverlapDefinition1}
    \mathcal{C}_{S,S'} \equiv 
    \sum_{l}
    \bigg|\langle S'|\hat{n}_{l,\vex{q}}|S\rangle\bigg|^2.
\end{align}
In $\mathcal{C}_{S,S'}$, the wavefunction overlap is given by
\begin{align}
\label{OverlapDefinition2}
    \langle S'|\hat{n}_{l,\vex{q}}|S\rangle = \sum_{\sigma,\vex{G},\vex{G'}}
    V^*_{b',\vex{k'}; \sigma, l, \vex{G}}
    V_{b,\vex{k}; \sigma, l, \vex{G'}}
    \delta_{\vex{k'}+\vex{G'},\vex{k}+\vex{G}+\vex{q}},
\end{align}
where $\vex{k},\vex{k}'$ take values in the (moir\'{e}-)Brillouin zone and $\vex{G},\vex{G}'$ take values on the (moir\'{e}) reciprocal lattice. The matrix elements $V$ define the (moir\'{e}-)Bloch wave functions that are eigenstates of the Hamiltonian:
\begin{align}
\label{MoireBloch}
\tilde{c}^\dagger_{s, \xi, b,\vex{k}} = \sum_{\sigma, l, \vex{G}} V_{b,\vex{k}; \sigma, l, \vex{G}}
\ c^\dagger_{s,\xi,\sigma,l,\vex{k}+\vex{G}},
\end{align}
such that
\begin{align}
    H^e = 
    \sum_{s,\xi,b,\vex{k}}
    \e_{s,\xi,b,\vex{k}}
    \tilde{c}^\dagger_{s, \xi, b,\vex{k}}
    \tilde{c}_{s, \xi, b,\vex{k}}.
\end{align}

%%%%%%%%%%%%%%%%%%%%%%%%%%%%%%%%%%%%%%%%%%%%%%%%%%%%%%%%%
%%%%%%%%%%%%%%%%%%%%%%%%%%%%%%%%%%%%%%%%%%%%%%%%%%%%%%%%%
%%%%%%%%%%%%%%%%%%%%%%%%%%%%%%%%%%%%%%%%%%%%%%%%%%%%%%%%%
% Ultrasubsonic transport theory: derivation details
%%%%%%%%%%%%%%%%%%%%%%%%%%%%%%%%%%%%%%%%%%%%%%%%%%%%%%%%%
%%%%%%%%%%%%%%%%%%%%%%%%%%%%%%%%%%%%%%%%%%%%%%%%%%%%%%%%%
%%%%%%%%%%%%%%%%%%%%%%%%%%%%%%%%%%%%%%%%%%%%%%%%%%%%%%%%%

\section{III.\ Ultrasubsonic transport theory: derivation details}
We provide additional supporting details for the derivation of the main results of ultrasubsonic (USS) transport theory, Eqs.~(6-7), and state the general version of the particle-hole-symmetric result given in Eq.~(7). As discussed in the main text, the technical advantage of the USS limit is the justification of the approximations $\mathcal{F}^{\mu,T}(S,S') \approx \mathcal{F}^{\mu,T}(S,\bar{S})$ and $\tau(S') \approx \tau(\bar{S})$ which replace the relaxation times and the thermal occupancy function at points along the scattering manifold ($S'$) by their values at the compliment point ($\bar{S}$). Applying these simplifications to Eq.~(\ref{RelaxationLengthSelfConsistency}) [Eq.~(2) in the main text], we have
\begin{align}
\label{DerivationRootEquation}
    1 &= 
    \frac{\pi D^2}{\hbar \rho_M v_p^2}
     \mathcal{F}^{\mu,T}_{S,\bar{S}} 
    \left(
    \tau_{S}
    \left[
    \int
    \frac{d^2\vex{k}}{(2\pi)^2}
    \tilde{\Delta}_{S,S'}
    \tilde{F}_{S,S'}
    \right]
    - 
    \tau_{\bar{S}}
    \left[
    \frac{1}{|\vex{v}_{S}|}
    \int
    \frac{d^2\vex{k}}{(2\pi)^2}
    \tilde{\Delta}_{S,S'}
    \mathcal{C}_{S,S'}
    |\vex{v}_{S'}|
    \cos\theta_{\vex{v}}
    \right]
    \right),
    \\
\label{DerivationRootEquationX}
    &\equiv
    \frac{D^2}{2 \rho_M \hbar^3 v_p^4}
    |\e-\bar{\e}| 
    \mathcal{F}^{\mu,T}_{S,\bar{S}}
    \left[
    X_S \tau_{S} 
    - 
    \tilde{X}_S \tau_{\bar{S}}
    \right],
\end{align}
where we have defined the $(T,\mu)$-independent factors
\begin{align}
\label{XsDefinition}
    X_S 
    &\equiv
    \frac{2\pi (\hbar v_p)^2}{|\e - \bar{\e}|}
    \int
    \frac{d^2\vex{k}}{(2\pi)^2}
    \tilde{\Delta}_{S,S'}
    \mathcal{C}_{S,S'},
    \\
\label{tildeXsDefinition}    
    \tilde{X}_S 
    &\equiv
    \frac{2\pi (\hbar v_p)^2}{|\e - \bar{\e}|}
    \int
    \frac{d^2\vex{k}}{(2\pi)^2}
    \tilde{\Delta}_{S,S'}
    \mathcal{C}_{S,S'}
    \frac{v_{s'}}{v_s}
    \cos{\theta_{v,v'}},
\end{align}
which encode wavefunction overlap and band geometry data. To motivate the prefactors in the definitions above, we note that if we approximate the scattering manifold as a circle around $\bar{S}$, we simply have
\begin{align}
\label{XsUltraApprox2}
    X_S 
    &\approx
    \frac{2\pi (\hbar v_p)^2}{|\e - \bar{\e}|}
    \int
    \frac{dq\ q}{(2\pi)}
    \delta[|\e-\bar{\e}| - \hbar v_p q]
    \int_0^{2\pi}
    \frac{d\phi}{(2\pi)}
    \mathcal{C}_{S,S'}
    = 
    \int_0^{2\pi}
    \frac{d\phi}{(2\pi)}
    \mathcal{C}_{S,S'}.
\end{align}
We emphasize that $X, \tilde{X}$ may depend heavily on the state index $S$ via the wavefunction. 

In Eq.~(\ref{DerivationRootEquationX}), we have recovered Eq.~(5) in the main text. Pairing this equation with the corresponding one for the state $\bar{S}$ gives a 2x2 matrix equation:
\begin{align}
\label{SelfConsistentMatrixEquation}
    \frac{D^2}{2 \rho_M \hbar^3 v_p^4}
    |\e-\bar{\e}| 
    \begin{bmatrix}
    \mathcal{F}^{\mu,T}_{S,\bar{S}} X_S & -\mathcal{F}^{\mu,T}_{S,\bar{S}} \tilde{X}_S  \\
    -\mathcal{F}^{\mu,T}_{\bar{S},S} \tilde{X}_{\bar{S}} & \mathcal{F}^{\mu,T}_{\bar{S},S} X_{\bar{S}}
    \end{bmatrix}
    \begin{bmatrix}
    \tau_S \\
    \tau_{\bar{S}}
    \end{bmatrix}
    &= 
    \begin{bmatrix}
    1 \\
    1
    \end{bmatrix}.
\end{align}
Simply inverting the 2x2 matrix and simplifying gives the general solution for the relaxation time stated in Eq.~(6) of the main text. Combining the relaxation time formula [Eq.~(6)] with the resistivity formula [Eq.~(1,\ref{ResistivityDefinition})] gives the master formula for the resistivity of the ultra-subsonic fermion system:
\begin{align}
\label{GeneralResistivityMasterFormula}
    \frac{\delta^{ij}}{\rho}
    &=
    \frac{e^2 \rho_M (\hbar v_p)^4}{\hbar D^2 k_BT}
    \frac{1}{\mathcal{A}}
    \sum_{\e_S > 0}
    \frac{1}{|\e-\bar{\e}|^2}
    \frac{
    \sinh\left(\frac{|\e-\bar{\e}|}{2k_B T}\right)
    }
    {
    \cosh\left(\frac{\bar{\e}-\mu}{2k_B T}\right)
    \cosh\left(\frac{\e-\mu}{2k_B T}\right)
    }
    \frac{1}
    {X_S X_{\bar{S}} - \tilde{X}_S\tilde{X}_{\bar{S}}}
    \left\lgroup
    \begin{aligned}
    &(\tilde{X}_S + X_{\bar{S}})v_S^iv_S^j + (\tilde{X}_{\bar{S}} + X_S)v_{\bar{S}}^iv_{\bar{S}}^j
    \\
    +&
    \sinh\left(\frac{\bar{\e}-\e}{2 k_B T}\right)
    \sinh\left(\frac{\bar{\e}+\e-2\mu}{2 k_B T}\right)
    \\
    &\times
    \left[
    \frac{
    X_{\bar{S}}v_S^iv_S^j
    }
    {
    \cosh\left(\frac{\e-\mu}{2k_B T}\right)^2
    }
    -
    \frac{
    X_Sv_{\bar{S}}^iv_{\bar{S}}^j
    }
    {\cosh\left(\frac{\bar{\e}-\mu}{2k_B T}\right)^2}
    \right]
    \end{aligned}
    \right\rgroup.
\end{align}
We emphasize that the complexity of the particle-hole-symmetry breaking terms in the brackets of Eq.~(\ref{GeneralResistivityMasterFormula}) significantly complicated the competition between the regimes of linear-in-$T$ resistivity and perfect conductivity discussed in the main text. In the case of a particle-hole (PH) symmetric system, the formulae for the USS relaxation time simplifies to
\begin{align}
\label{PHSymmetricTau}
    \tau_S 
    &=
    \frac{\rho_M (\hbar v_p)^4}{\hbar D^2} 
    \frac{1}{|\e|^2}
    \sinh\left(\frac{|\e|}{k_B T}\right)
    \frac{1}{X^2_S - \tilde{X}^2_S}
    \left[
    X_S
    \frac{\cosh\left(\frac{\e+\mu}{2k_B T}\right)}{\cosh\left(\frac{\e-\mu}{2k_B T}\right)}
    +
    \tilde{X}_S
    \frac{\cosh\left(\frac{\e-\mu}{2k_B T}\right)}
    {\cosh\left(\frac{\e+\mu}{2k_B T}\right)}
    \right].
\end{align}
Inserting Eq.~(\ref{PHSymmetricTau}) into the resistivity formula [Eqs.~(1,\ref{ResistivityDefinition})] gives the PH-symmetric resistivity master formula [Eq.~(7)] analyzed in the main text.

As noted in the main text, the high-$T$ limit ($k_B T \gg |\e-\bar{\e}|$) gives the expected $\tau \propto T^{-1}$ scattering rate of the equipartition regime [Eq.~(3)]. In the case of USS scattering, we see that the constants $c_S$ are given explicitly by
\begin{align}
    c_S = 
    \frac{2 \rho_M (\hbar v_p)^4}{\hbar D^2}
    \frac{1}{|\e-\bar{\e}|}
    \frac{X_{\bar{S}} + \tilde{X}_S}{X_S X_{\bar{S}} - \tilde{X}_S \tilde{X}_{\bar{S}}}.
\end{align}

%%%%%%%%%%%%%%%%%%%%%%%%%%%%%%%%%%%%%%%%%%%%%%%%%%%%%%%%%
%%%%%%%%%%%%%%%%%%%%%%%%%%%%%%%%%%%%%%%%%%%%%%%%%%%%%%%%%
%%%%%%%%%%%%%%%%%%%%%%%%%%%%%%%%%%%%%%%%%%%%%%%%%%%%%%%%%
% Hierarchy of approximations in ultrasubsonic limit
%%%%%%%%%%%%%%%%%%%%%%%%%%%%%%%%%%%%%%%%%%%%%%%%%%%%%%%%%
%%%%%%%%%%%%%%%%%%%%%%%%%%%%%%%%%%%%%%%%%%%%%%%%%%%%%%%%%
%%%%%%%%%%%%%%%%%%%%%%%%%%%%%%%%%%%%%%%%%%%%%%%%%%%%%%%%%

\section{IV.\ Hierarchy of approximations in ultrasubsonic limit}

We formalize the hierarchy of approximations that we collectively refer to as ``ultra subsonic" (USS):
\begin{enumerate}
    \item ``USS 1'': $\tau(S') \approx \tau(\bar{S})$ along the scattering manifold.
    \item ``USS 2'': In addition to USS 1, we further assume that $\mathcal{F}^{\mu, T}_{S,S'} \approx \mathcal{F}^{\mu, T}_{S,\bar{S}}$ along the scattering manifold.
    \item ``USS 3'': In addition to USS 1 and USS 2, we also approximate the scattering manifold as a small circle around the point $\bar{S}$, as in Eq.~(\ref{XsUltraApprox2}).
\end{enumerate}
We note that USS 1 enables the reduction of the self-consistent integra equation for the relaxation times [Eqs.~(2,\ref{RelaxationLengthSelfConsistency})] to a simple matrix equation admitting closed form solution. Technically the final ``master" formula for ultra subsonic scattering [Eq.~(\ref{GeneralResistivityMasterFormula})] below depends only on USS 1. However, USS 2 is necessary to remove the dependence of $T,\mu$ from $X_S, \tilde{X}_S$ and make the thermal scaling of the resistivity transparent. The approximation USS 3 is not necessary for our main results but is of utility in the explicit evaluation of the factors $X_{S},\tilde{X}_S$.

%%%%%%%%%%%%%%%%%%%%%%%%%%%%%%%%%%%%%%%%%%%%%%%%%%%%%%%%%
%%%%%%%%%%%%%%%%%%%%%%%%%%%%%%%%%%%%%%%%%%%%%%%%%%%%%%%%%
%%%%%%%%%%%%%%%%%%%%%%%%%%%%%%%%%%%%%%%%%%%%%%%%%%%%%%%%%
% Ultrasubsonic Dirac cone
%%%%%%%%%%%%%%%%%%%%%%%%%%%%%%%%%%%%%%%%%%%%%%%%%%%%%%%%%
%%%%%%%%%%%%%%%%%%%%%%%%%%%%%%%%%%%%%%%%%%%%%%%%%%%%%%%%%
%%%%%%%%%%%%%%%%%%%%%%%%%%%%%%%%%%%%%%%%%%%%%%%%%%%%%%%%%

\section{V.\ Ultrasubsonic Dirac cone}

As a concrete example, we apply our results to the ultrasubsonic Dirac cone, where the simple band structure can be used to to evaluate the formulae explicitly. We must evaluate $X_{S}$ and $\tilde{X}_S$. By symmetry, $X_{S}$ and $\tilde{X}_S$ will only depend on the state $S$ via the energy $\e$, though we will see that in the case of the Dirac cone they are also $\e$-independent. We let $r \equiv v_F/v_p$. 

For the subsonic Dirac cone, we have $\cos\theta_{v} = -\cos\theta$, where $\theta$ is the momentum angle and 
\begin{align}
    \mathcal{C}_{S,S'} = \frac{1}{2}(1-\cos\theta).
\end{align}
Analyzing the kinematics of phonon scattering on the Dirac cone, we see that the scattering manifold for the interband transitions in the subsonic Dirac cone is an ellipse parameterized by the momentum transfer vector 
\begin{align}
    q(\phi) 
    = 
    \frac{2rk}{1-r^2}
    (1+r\cos\phi),
\end{align}
where $\phi$ is the angle between the momentum of the initial fermion state and the phonon state. Some trigonometry gives the momentum angle ($\theta$) in terms of $\phi$:
\begin{align}
    \cos\theta 
    =
    1
    -
    \frac{2 r^2 \sin^2\phi}{1 + r^2 + 2 r \cos\phi}.
\end{align}
We may then evaluate
\begin{align}
    X_S(r) 
    &=
    \frac{2\pi (\hbar v_p)^2}{|\e - \bar{\e}|}
    \int
    \frac{dq\ q}{(2\pi)}
    \int_0^{2\pi}
    \frac{d\phi}{(2\pi)}
    \delta[|\e-\e'| - \hbar v_p q]
    \mathcal{C}_{S,S'},
    \\
    &=
    \frac{\hbar v_p}{4 |\e|}
    \int_0^{2\pi}
    \frac{d\phi}{2\pi}
    q(\phi)
    \big(1-\cos[\theta(\phi)]\big),
    \\
    &=
\label{ExactFormulaX}     
    \frac{1}{2}
    \int_0^{2\pi}
    \frac{d\phi}{2\pi}
    \frac{1+r\cos\phi}{1-r^2}
    \left[
    \frac{2 r^2 \sin^2\phi}{1 + r^2 + 2 r \cos\phi}
    \right].
\end{align}
We note that Eq.~(\ref{ExactFormulaX}) is an exact evaluation of $X_S$ that does not make use of the approximation ``USS 3", and we point out that the final answer has no dependence on the state $S$. Similarly,
\begin{align}
    \tilde{X}_S(r) 
    &\approx
    \frac{-1}{2}
    \int_0^{2\pi}
    \frac{d\phi}{2\pi}
    \frac{1+r\cos\phi}{1-r^2}
    \left[
    \frac{2 r^2 \sin^2\phi}{1 + r^2 + 2 r \cos\phi}
    \right]
    \left[
    1
    -
    \frac{2 r^2 \sin^2\phi}{1 + r^2 + 2 r \cos\phi}
    \right],
    \\
    &=
\label{ExactFormulaXTilde}    
    -X(r)
    +
    \frac{1}{2}
    \int_0^{2\pi}
    \frac{d\phi}{2\pi}
    \frac{1+r\cos\phi}{1-r^2}
    \left[
    \frac{2 r^2 \sin^2\phi}{1 + r^2 + 2 r \cos\phi}
    \right]^2.
\end{align}
Like Eq.~(\ref{ExactFormulaX}) for $X(r)$, Eq.~(\ref{ExactFormulaXTilde}) gives an exact expression for $\tilde{X}_S$ and shows that there is no dependence on the initial state $S$.

Working to leading order in $v_F/v_p$ we simply have
\begin{align}
    X = \frac{1}{2}\left(\frac{v_F}{v_p}\right)^2,\ \ \ \tilde{X} = \frac{3}{4}\left(\frac{v_F}{v_p}\right)^4 - X.
\end{align}
The above results show that 
\begin{align}
\label{XRatio1DiracCone}
    \frac{1}{X_S - \tilde{X}_S} 
    &\approx 
    \frac{1}{r^2}\frac{4}{4-3r^2}
    = 
    \left(\frac{v_p}{v_F}\right)^2
    \frac{4v_p^2}{4 v_p^2 - 3 v_F^2}
    \\
\label{XRatio2DiracCone}    
    \frac{X_S}{X^2_S - \tilde{X}^2_S} 
    &\approx  
    \frac{2}{3r^4}\frac{4}{4-3r^2}
    =
    \frac{2}{3}
    \left(\frac{v_p}{v_F}\right)^4
    \frac{4v^2_p}{4v^2_p - 3v^2_F}.
\end{align}
Applying Eqs.~(\ref{XRatio1DiracCone}-\ref{XRatio2DiracCone}) to the PH-symmetric resistivity formula [Eq.~(7)] and using the Dirac cone band structure gives a concrete formula for the resistivity of the USS Dirac cone:
\begin{align}
    \frac{1}{\rho} 
    &\approx
    \frac{1}{k_B T}
    \frac{2}{\pi}
    \frac{e^2 \rho_M \hbar v_p^4}{ D^2}
    \left(\frac{v_p}{v_F}\right)^2
    \frac{v_p^2}{4v_p^2 - 3v_F^2}
    \left\lgroup
    \begin{aligned}
    &
    \int_0^{{\Lambda}}
    \frac{d\e}{\e}
    \frac{\sinh\left(\frac{\e}{k_B T}\right)}
    {\cosh\left(\frac{\e+\mu}{2k_B T}\right)\cosh\left(\frac{\e-\mu}{2k_B T}\right)}
    \\
    +
    &
    \frac{1}{3}
    \left(\frac{v_p}{v_F}\right)^2
    \sinh\left(\frac{|\mu|}{k_B T}\right)^2
    \int_0^{{\Lambda}}
    \frac{d\e}{\e}
    \left[
    \frac{
    \sinh\left(\frac{\e}{k_B T}\right)
    }
    {
    \cosh\left(\frac{\e+\mu}{2k_B T}\right)
    \cosh\left(\frac{\e-\mu}{2k_B T}\right)
    }
    \right]^3
    \end{aligned}
    \right\rgroup.
\end{align}
Specializing to the $\mu = 0$ case, we find
\begin{align}
\label{HalfFilingUSSDiracCone}
    \rho
    &\approx
    \frac{\pi k_B T D^2}{e^2 \rho_M \hbar v_p^4}
    \frac{v_F^2}{v_p^2}
    \left(1-\frac{3}{4}\frac{v_F^2}{v_p^2}\right)
    \log\left[
    \frac{\Lambda}{k_B T}
    \right]^{-1}
\end{align}
which gives a purely linear in $T$ resistivity, up to cutoff-dependent logarithmic corrections. Away from $\mu = 0$ but in the extreme low-$T$ limit, we may make the simplification from Eq.~(9) of the main text. This gives
\begin{align}
    \rho 
    \approx 
    \frac{v_F^4}{v_p^8}
    \frac{\pi D^2 k_B T}{e^2 \rho_M \hbar} 
    \frac{1}{\log(\Lambda/|\mu|)}e^{-2|\mu|/(k_B T)}
\end{align}
in the $r,T \rightarrow 0$ asymptote. The final resistivity formulae for the USS Dirac cone demonstrate the general features of USS transport, showing robust linear-in-$T$ resistivity partially generated by states far from the Fermi level and a crossover to perfect conductivity with exponentially suppressed scattering when $k_B T \ll 2|\mu|$. We note that the prefactor contains nontrivial power-law dependencies on the velocities $v_F$ and $v_p$ that could prove useful for experimental probes of USS transport physics in systems for which the Dirac cone is a useful model.

%%%%%%%%%%%%%%%%%%%%%%%%%%%%%%%%%%%%%%%%%%%%%%%%%%%%%%%%%
%%%%%%%%%%%%%%%%%%%%%%%%%%%%%%%%%%%%%%%%%%%%%%%%%%%%%%%%%
%%%%%%%%%%%%%%%%%%%%%%%%%%%%%%%%%%%%%%%%%%%%%%%%%%%%%%%%%
% Ultrasubsonic fermions in MATBLG
%%%%%%%%%%%%%%%%%%%%%%%%%%%%%%%%%%%%%%%%%%%%%%%%%%%%%%%%%
%%%%%%%%%%%%%%%%%%%%%%%%%%%%%%%%%%%%%%%%%%%%%%%%%%%%%%%%%
%%%%%%%%%%%%%%%%%%%%%%%%%%%%%%%%%%%%%%%%%%%%%%%%%%%%%%%%%

\section{VI.\ Ultrasubsonic fermions in MATBLG}
\begin{figure}[t!]
\includegraphics[angle=0,width=.9\textwidth]{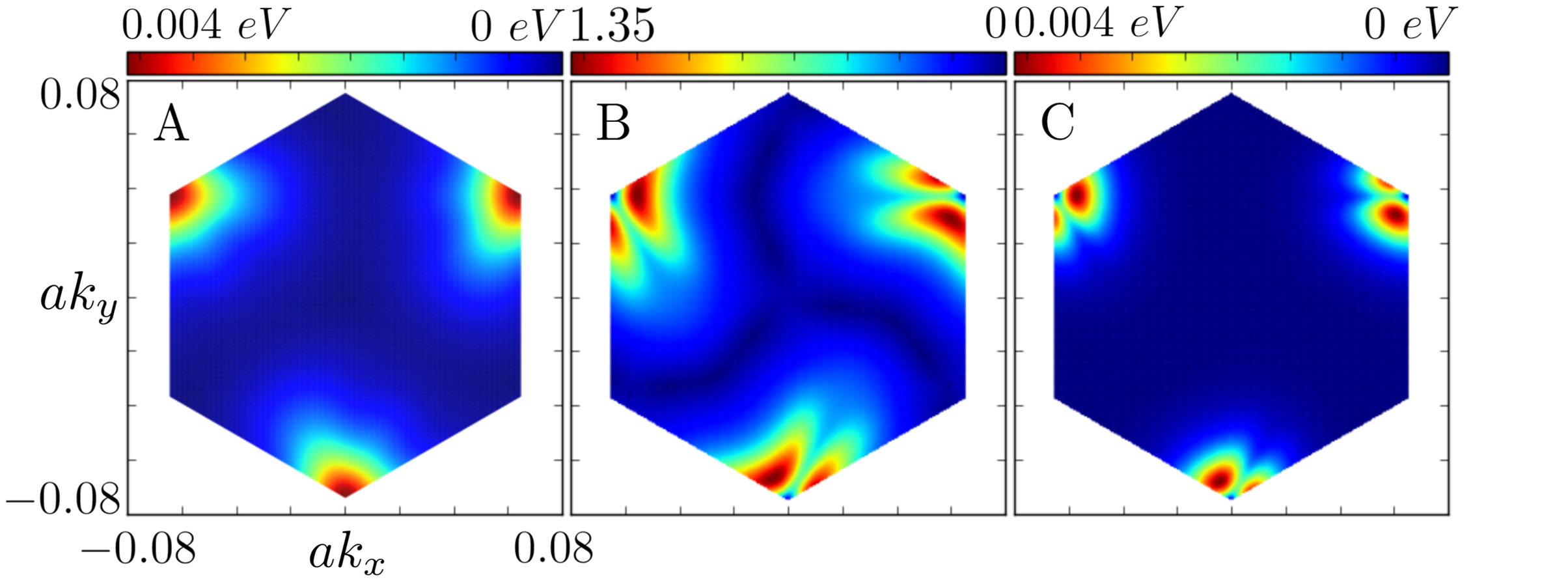}
\includegraphics[angle=0,width=.9\textwidth]{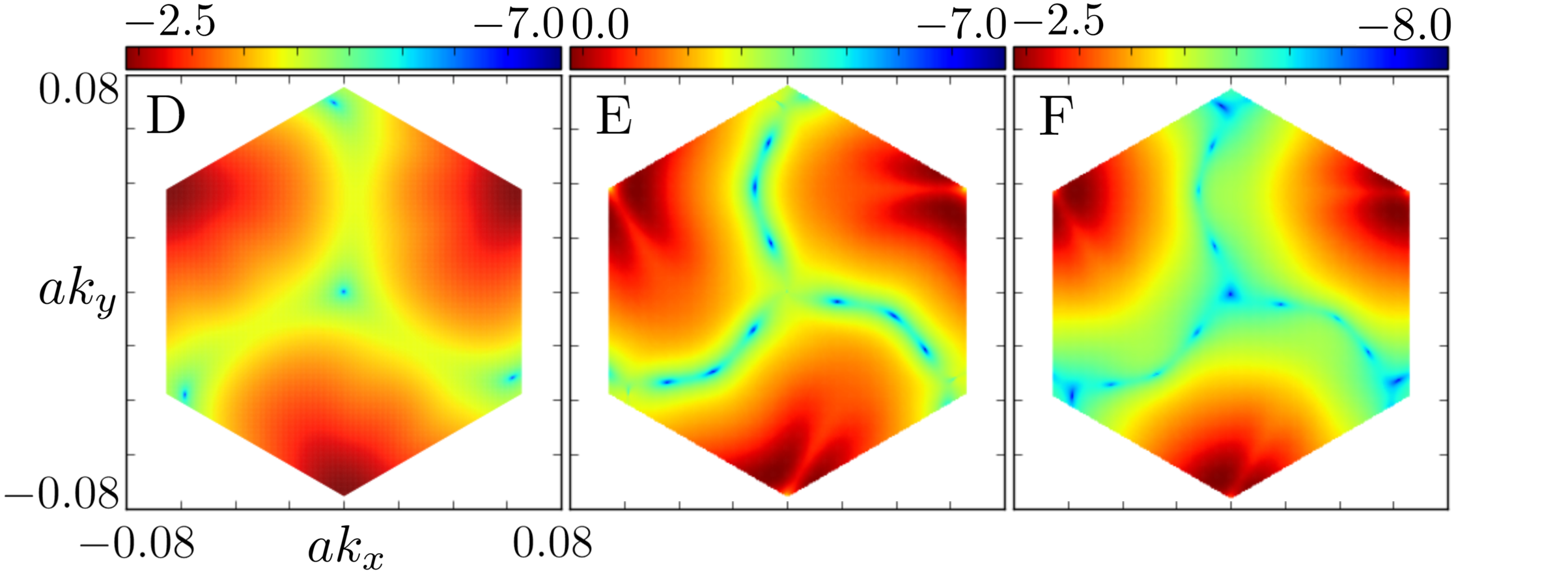}

\caption{We demonstrate the extent of the applicability of the ultrasubsonic scattering theory to MATBLG by providing numerical heat maps of some crucial quantities, calculated for the Bistritzer MacDonald model for the non-interacting band structure of MATBLG. We use the standard BM Hamiltonian with the interlayer hopping parameters $\omega_0 = 90\ meV$, $\omega_1 = 117\ meV$ and the bare graphene Dirac cone velocity $v_F = 10^6 m/s$, which places the ``magic angle" at $\theta = 1.025^{\circ}$. We take $v_p = 2 \cdot 10^4 m/s$, consistent with phonon speeds in single-layer graphene \cite{Wu2019_phonon}. In the left column (A,D), we depict the energy difference between the two NFBs, $|\e_S - \e_{\bar{S}}|.$ We note that it is very small except for a single peak. In the center (C,E), we calculate the dimensionless ratio $v_F/v_p$.  We see that the majority of the band is strongly subsonic, though there are supersonic Bloch states near the peak. Finally, in the right column (C,F), we plot the product of these two quantities, which is the energy scale justifying the USS approximations that enable our theory. For each quantity, we plot the data in a linear scale on the top row (A,B,C) and a log scale on the bottom row (D,E,F).}
\label{BisritzerMacDonaldModel}
\end{figure}

We close with a more quantitative discussion of the applicability of USS transport theory to MATBLG in particular. In Fig.~\ref{BisritzerMacDonaldModel}, using the Bistritzer-MacDonald (BM) model for the non-interacting bandstructure of MATBLG \cite{Bistritzer_2011}, we give numerical heat maps of the interband spacing, the ratio $v_F/v_p$, and the product of these quantities for the NFBs. It is clear from these figures that the vast majority of the Bloch states in the moir\'{e} Brillouin zone are very well captured by the subsonic limit, with energy spacing below $0.0005\ eV$ and $v_F/v_p < 0.01$. For some states, these values approach zero. We further emphasize that while these systems due have significant twist-angle disorder, they are almost entirely free from impurities that would cause elastic scattering, indicating that transport is indeed likely phonon-dominated.

Even in the non-interacting BM model, the NFBs are not entirely subsonic - there is a small ``high"-energy hump in the dispersion characterized by energy separations a large as $0.004\ eV$ and $v_F/v_p$ as large as $1.35$. The coexistence of subsonic and supersonic states in the band would need to be taken into account to correctly apply USS scattering theory to the BM band structure in a quantitative way. For example, some subsonic states near the high-energy region may still experience intraband scattering. As discussed in the paper, it is not clear that a direct application of USS theory to the BM band will give quantitatively accurate transport predictions at low-$T$, due to possible band renormalizations. Nevertheless, even after significant band renormalization, it is possible that large portions of the NFB are well-captured by the ultrasubsonic limit and exhibit the related low-$T$ transport phenomena.

%%%%%%%%%%%%%%%%%%%%%%%%%%%%%%%%%%%%%%%%%%%%%%%%%%%%%%%%%
%%%%%%%%%%%%%%%%%%%%%%%%%%%%%%%%%%%%%%%%%%%%%%%%%%%%%%%%%
%%%%%%%%%%%%%%%%%%%%%%%%%%%%%%%%%%%%%%%%%%%%%%%%%%%%%%%%%
% Relaxation times in Boltzmann kinetic theory
%%%%%%%%%%%%%%%%%%%%%%%%%%%%%%%%%%%%%%%%%%%%%%%%%%%%%%%%%
%%%%%%%%%%%%%%%%%%%%%%%%%%%%%%%%%%%%%%%%%%%%%%%%%%%%%%%%%
%%%%%%%%%%%%%%%%%%%%%%%%%%%%%%%%%%%%%%%%%%%%%%%%%%%%%%%%%

%\section{VII.\ Relaxation times in Boltzmann kinetic theory}

\bibliography{BIB}